\documentclass[letterpaper]{article}
\usepackage{aaai}
\usepackage{times}
\usepackage{helvet}
\usepackage{courier}
\nocopyright
\usepackage{tabularx, multirow, booktabs}
\usepackage{comment}
\usepackage{multirow}
\usepackage{mathtools}
\usepackage{diagbox}
\usepackage[normalem]{ulem}

\usepackage[algoruled, boxed, lined, noend]{algorithm2e}
\usepackage[font=footnotesize]{caption}
\usepackage[font=footnotesize]{subcaption}
\usepackage[utf8]{inputenc}
\usepackage{xcolor}

\newcommand\tab[1][1cm]{\hspace*{#1}}
\usepackage{amsfonts}
\usepackage{booktabs}
\usepackage{siunitx}
\newcolumntype{L}[1]{>{\raggedright\arraybackslash}p{#1}}
\newcolumntype{C}[1]{>{\centering\arraybackslash}p{#1}}
 
\usepackage{todonotes}

\begin{document}
%
\title{On synthetic data generation for anomaly detection in complex social networks}
 \author{
{\bf \Large Andreea Sistrunk,\textsuperscript{1,2} 
Vanessa Cedeno,\textsuperscript{1,3} 
Subhodip Biswas\textsuperscript{1}
}
\\  
\textsuperscript{1} Virginia Tech,  Department of Computer Science, 
\textsuperscript{3} Escuela Superior Politecnica del Litoral, Guayaquil, Ecuador\\
\textsuperscript{2} U.S. Army - Engineering, Research, and Development Center - Geospatial Research Laboratory}

\maketitle
\begin{abstract}
This paper studies the feasibility of synthetic data generation for mission-critical applications. The emphasis is on synthetic data generation for anomalous detection in complex social networks. In particular, the development of a heuristic generative model, capable of creating data for anomalous rare activities in complex social networks is sought. To this end, lessons from social and political literature are applied to prototype a novel implementation of the Agent-based Modeling (ABM) framework, based on simple social interactions between agents, for synthetic data generation in the context of terrorist profile desegregation. The conclusion offers directions for further verification, fine-tuning, and proposes future directions of work for the ABM prototype, as a complex-societal approach to synthetic data generation, by identifying heuristic hyper-parameter tuning methodologies to further ensure the generated data distribution is similar to the true distribution of the original data-sets. While a rigorous mathematical optimization for reducing the distances in distributions is not offered in this work, we opine that this prototype of an autonomous-agent generative complex social model is useful for studying and researching on pattern of life and anomaly detection where there is strict limitation or lack of sufficient data for using practical machine learning solutions in mission-critical applications.
\end{abstract}

\section{Introduction}

Despite consistent work towards lowering of crime rates, societies remain fragile and vulnerable to crime.
Knowledge of past events enforced by spatial coordinates of crime showed to be vital in rare events such as disaster and crime management \cite{malleson2015impact}. In consequence, many attempts have been made to understand crime, predict its patterns, as well as determine possible profiles of suspected people. Research of this network of relationships showed consistent positive impact in efficient resource allocation (officers, vehicles, surveillance cameras, etc.) and concrete results \cite{perry2013predictive}.
However, recorded information on crime is highly variable depending on type of crime and location. Additionally, grave or organized crime, such as terrorism, is a sporadic occurrence.

\begin{figure}[!htp]
     \begin{subfigure}[b]{0.45\linewidth}
         \centering
         \includegraphics[width=\textwidth, keepaspectratio]{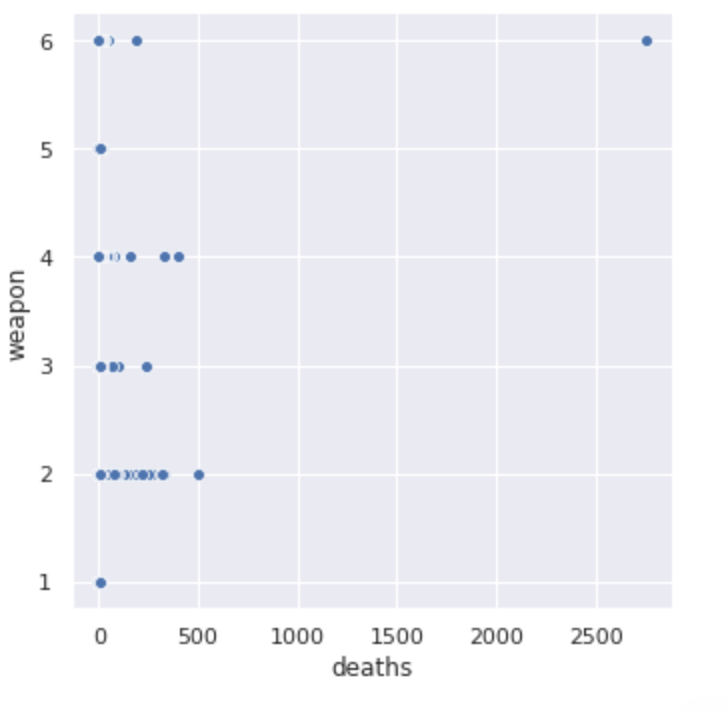}
         \caption{}
     \end{subfigure}
    ~
     \begin{subfigure}[b]{0.54\linewidth}
         \centering
         \includegraphics[width=\textwidth, keepaspectratio]{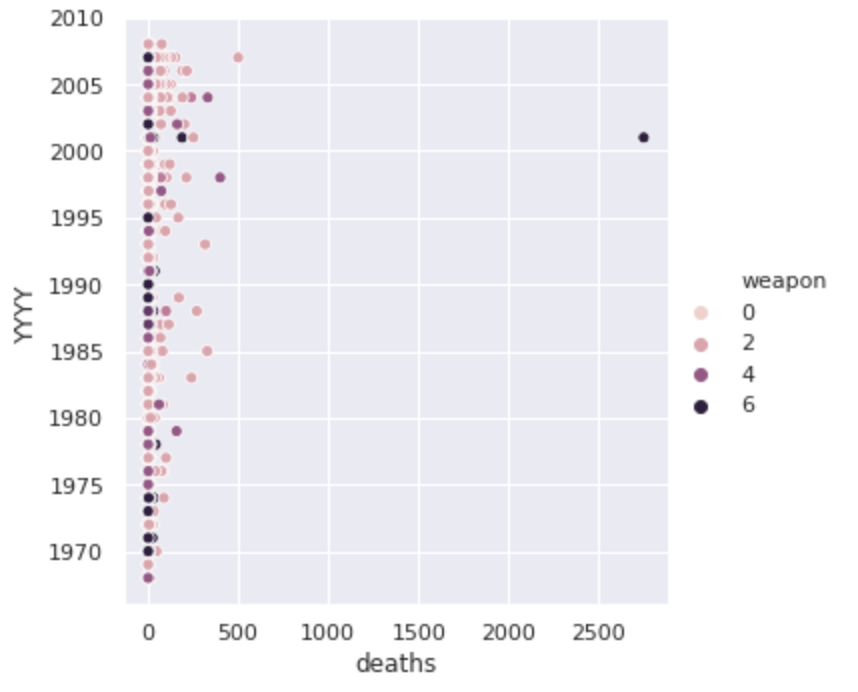}
         \caption{}
     \end{subfigure}

     \begin{subfigure}[b]{0.49\linewidth}
         \centering
         \includegraphics[width=\textwidth, keepaspectratio]{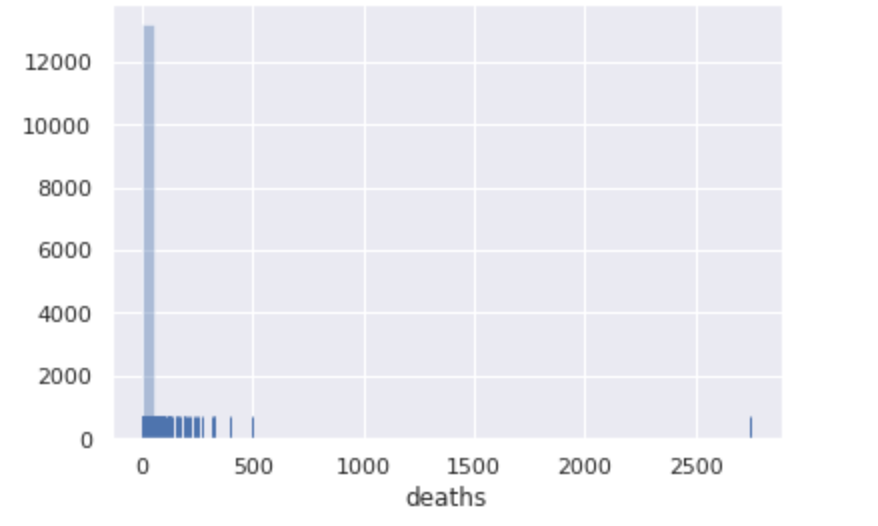}
         \caption{}
     \end{subfigure}
    ~
     \begin{subfigure}[b]{0.48\linewidth}
         \includegraphics[width=\textwidth, keepaspectratio]{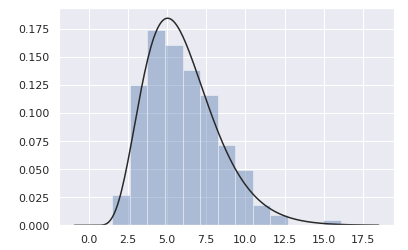}
         \caption{}
     \end{subfigure}
         \caption{RAND-MIPT: \textbf{(a)} deaths per weapon-type: 1 = biological/chemical agents, 2 = explosives or remote detonated explosives, 3 = fire/firebomb or arson, 4 = firearms, 5 = knives or sharp objects, 6 = other or unknown.
        \textbf{(b) } \textit{x-axis:} number of deaths per attack, \textit{y-axis:} year of recorded incident. The legend shows recorded weapon type. 
       \textbf{(c)} \textit{x-axis:} number of deaths, \textit{y-axis:} number of incidents. Most of the incidents recorded have no (zero) casualties (deaths), one recorded event has over 2500 deaths.
        \textbf{(d)} kernel density - gamma distribution fit on death tolls. }
        \label{fig:difficulty}
    \end{figure}

In most countries, terrorism accounts for an average of $0.01$ \% of the total deaths Figure~\ref{fig:difficulty}~\cite{ritchiehasellappelroser2013}. Yet every attack causes a serious negative psychological impact on society and constitutes significant disruptions. For example, the 9/11 attacks costed the US over 100 billion dollars and terrorist attacks between 2004-2016 cost the European Union over 180 billion Euros~\cite{RAND2018}. Our limited understanding of such events comes in part, from them being a \textit{rare event problem}. 
Its erratic occurrence and non-traditional crime triggers add to the fear it instills.

In consequence, recorded data are inherently insufficient for quantitative studies, essential in the modeling, establishing reliable patterns, and generating analysis for prevention and early conflict mitigation. As with other instances where we do not have sufficient collected data, scientists are used to augmenting with synthetic data, following main parameters and features. 

Producing synthetic data that exhibits similar threats as real data would be useful to augment collected data points. It could provide an \textit{additional avenue to study sensitive information, otherwise unavailable due to scarcity or classification}. Subsequently the ability to produce synthetic data that follows the characteristics of recorded data would allow utilizing modern Artificial Intelligence (AI) techniques, mostly unavailable to this point for this problem. 


To produce synthetic data in the absence of an established distribution, scientists could employ the Gaussian distribution. However, this distribution would not produce long tails, typically associated with rare events, thus not capturing rare events happening within three ${\sigma}$ standard deviations 
\cite{galarnyk2019}. Therefore traditional methodology for synthetic data generation does not apply to this problem. 

\noindent
\subsection{Background and Motivation}  
\textit{Crime vs Organized Crime:}  In terms of fatalities and property loss, all crime has similar consequences. More so, records about people convicted of terrorism include other criminal activities \cite{mullins2009parallels}. Yet small fundamental distinguishing differences between other types of crime, about which we have more data, and terrorist activities, in which recorded data is limited, make the more restrictive samples unsuitable for transfer knowledge, because of the crime motive.

In most types of crime, the criminal seeks a personal benefit of some sort; however terrorists often times are serving an ideology and consider their acts as personal altruist sacrifice \cite{reeve2020islamist}. 
Initially our team sought a transfer knowledge methodology between terrorism and other areas of crime in which more data has been captured. Yet because many terrorist acts are committed in the name of altruism, the lessons we learned in predictive policing (an area of law enforcement that uses data from disparate sources and predictive analytic techniques to gather information and use for crime prevention and increased efficiency of response to future crime) do not transfer to the study of terrorism.

\textit{Data Sets Investigated:} (i) Publicly available data offered under the Global Database of Events, Language, and Tone. (GDELT) Project contains news gathered daily since 1979; (ii) The University of Maryland Study of Terrorism and Responses to Terrorism (START) database; (iii) National Security Research Division (RAND) database about worldwide terrorism incidents and particularly the RAND-MIPT terrorism incident database; (iv) Center for International Security and Corporation (CISAC) from Stanford for mapping militants; (v) World Bank database for statistics.

Recorded data shows an incidence of a serious terrorist event (such as 9/11) with probability of 0.01\% in most countries and goes as high as 7\% in the terrorism epicenters of the planet (World Data statistics and RANDT-MIPT terrorist data). Figure~\ref{fig:difficulty} is a visual illustration of RAND-MIPT database of 13274 deadly events gathered since 1968 \cite{Clauset2012}. Based on this data, which is in line with other databases, the occurrence of an event such as 9/11 with 2,977 deaths and 6,000 injuries  it is truly a rare event. Figure 1(a) illustrates casualties per weapon type.  Figure~\ref{fig:difficulty}~(b) shows a distribution of deaths per year, colored by weapon type and Figure~\ref{fig:difficulty}~(c)  shows most terrorist attacks do not have any casualty. Over 12000 recorded attacks have zero deaths. Still, despite its rare incidence, over 50\% of the people surveyed in the US alone, live in fear of being a victim of a terrorist activity \cite{ritchiehasellappelroser2013}.  

\subsection{Background} With scarce data, most research in this area remains qualitative. The few quantitative works conducted by universities and science foundations draw similarities to other areas of science such as social networks \cite{ChangChenFomby2017}, or employ simulation such as ABM. 

  ABM has become a known technique to simulate complex systems otherwise difficult not only to predict, but even to describe. So in an area where descriptive or prescriptive theories alone do not transfer well, ABM uses a simple set of rules to describe a complex process. Among first ABM models to raise to prominence, was Thomas Schelling’s Segregation Model (1971). Its implementation went on to win the Nobel prize, and subsequently was automated in many applications \cite{singh2009schelling}. Today ABM it is used to simulate agent interaction in computational sociology \cite{macy2002factors}, game theory \cite{cedeno:2019}, knowledge-based agents \cite{tecuci1999integrated}, or  dynamics of the structure and response to adverse reactions \cite{li2015agent}, etc.; and has a proposed standard protocol and implementations \cite{grimm2006standard}.
 
 Publications in the area of simulation of terrorist events  look at the overall terrorist groups, their behavior and interactions with little differentiation between the agent roles  \cite{tsvetovat2002knowing}.  Other research evaluates strategic options \cite{keller2010dismantling} or models events with variant input parameters \cite{gilbert2019agent}. 

 In-depth work~\cite{ilachinski2012modelling} focused on modeling the entire network of terrorist as self-adaptive societies. This tool continuously feeds from a variety of sources and we can see it as the state of the art in modelling in the area. While it has very extensive formulations, its availability and ability to reproduce as open source is highly restrictive. 

In consequence, predicting terrorist behavior is a constant high visibility matter with no easy answers.  In addition to insufficient recorded events, we do not have readily available mathematical models. Yet given the fear it instills, threat detection remains a hard, but vital problem. ABM modelling holds the promise in the ability to create synthetic societies using simple rules, could be a highly useful solution both for data augmentation, allowing better modelling, and subsequently unclassified analysis of restrictive data sets.


\begin{table}[!htpt]
\scriptsize
\caption{Hypothesis of Terrorist Traits}
\label{table:1}
\begin{tabular}{ccccc}
\toprule
   
   {Role} &{Education}&{Marriage} &{Wealth}&{Religion Training}
\\ \hline

     {Perpetrator}  & {low}  & {single}  & {poor} & {neutral}  \
     \\{Leader} &{high} & {married} & {neutral}&{trained while young}\
     \\{Financier}&{neutral}& {married}&{wealthy} &{trained older}\\
     
\hline
\end{tabular}
\normalsize
\end{table}

\subsection{Contributions}
In this paper we present  empirical evidence of a potential solution to producing synthetic data that follows recorded distributions for terrorist events. We do this by 
creating adaptive agents and interaction rules, subsequently allowing the society to evolve and mature. 
 The novelty of our work is in modeling terrorist interaction based on desegregated terrorist profiles. We recognized terrorist activity as only one small part of the societal interactions, therefore propose a simulation of holistic societal interactions within a geographical region. 
 In this work we are basing the agent profiles on political science research to offer an interdisciplinary insight into the problem. To carry out a simulation of a modeled society in which terrorist activity exists, we follow research based evidence proven reliable in qualitative studies from social and political sciences~\cite{perliger2016gap}. The political research shades light on the importance of dis-aggregating terrorist profiles. We model this desegregation using mathematical distributions and computer simulations to create individual agents that interact in the resulting complex environment~\cite{macal2005tutorial}. To  the further nuance the initial criteria, data is extracted from different geographic regions, with known distinctive profiles, and applied as a subsequent rule.

\begin{table*}[!htbp]
\small
  \begin{center}
   \caption{Table of Definitions}
    \label{table:2}
    \begin{tabular}{L{1.5cm}L{15cm}}
      \toprule
    {$\textbf{Parameter}$}&{$\textbf{Description}$} \\
    \hline 
    \textit{n} & Total number of the agents in a simulation at \textbf{state \textit{i}}\\
    $\alpha$ & Agents: total number of agents at \textbf{state \textit{i} }present in in the synthetic environment: $ \alpha_1^\textit{i}, \alpha_2^\textit{i},\alpha_3^\textit{i},...\alpha_i^\textit{i}, .... \alpha_n^\textit{i} $\\
    $\tau$ & Features: each agent has 9 features. The tensor of all features characterize an agent at a certain state. For an agent $\alpha$ at state \textit{i} we write tensor: \begin{math}
            \alpha^i=\{ \tau_1,\tau_2,\tau_3,\tau_4, \tau_5,\tau_6^i, \tau_7^i,\tau_8^i, \tau_9^i\}
            \end{math}\\
    m&Total number of states: an agent $\alpha$ in final state will look like \begin{math}
            \alpha^m=\{ \tau_1,\tau_2,\tau_3,\tau_4, \tau_5,\tau_6^m, \tau_7^m,\tau_8^m, \tau_9^m\}
            \end{math}\\
    \textit{w}& {Agent feature values $\tau$ generated based on the characteristics of the modelled geographical region} \\
    $\mu$ & Continuous probability distribution/normal distribution mean\\
    \textit{s}& Scale parameter proportional with the standard deviation \\
 \hline
  \bottomrule
    \end{tabular}
  \end{center}
\end{table*}
The terrorist profile desegregation distinguishes various types of roles an agent assumes. 
Therefore, we desegregated the terrorist label based on agent role within the organization in perpetrator: the actor that carries out a terrorist attack, group leader: the mastermind behind the attack, and financier: who supplies funds to support terrorist activities. 

To model a holistic society, in addition to agents tasked with generating an attack we added other agents with different roles in preventing/stopping
, or by-standing 
activities related to terrorism. In here we define rules to simulate terrorism recruiting, leading, and carrying the attack. These activities take place in a real society in addition to other agents' activities that have no connection with terrorism such as police and civilians. 

While we provide only incipient mathematical measures to show the basic similarities between the produced synthetic data and recorded data-points for terrorist activity, our contribution resides in constructing a synthetic society that functions on principles presented in socio-political literature~\cite{perliger2016gap}; bridging the knowledge between other areas of science into computer science, in a vital area that needs quantitative studies; as well as creating a prototype for synthetic data generation for mission critical application.

\section{ Models and Algorithm Description}
\paragraph{Conceptual Profiles.} Due to the various activities required to carry out a terrorist attack, terrorists need to perform different roles. Research in political science  depicts the gap between participation and violence in terrorist activities~\cite{perliger2016gap}. The authors conclude a possible correlation between socioeconomic status, marital status, religious education, and opportunity, as defining factors for someone to turn to terrorist activities as well the role they would play within a terrorist organization. The research connects the characteristic of terrorists to the role they might play in an attack with wealth, education, and marital status as key indicators for terrorists to become a perpetrator, leader, or financier. 
 For clarity we attempt to distinguish the characteristics of these individuals in our simulation separately.

The perpetrator plays a key role in carrying out the attack and in most cases their life is just one of the assumed sacrifices their role requires. These people are likely to be less educated, unemployed, are not married, and often local to the area~\cite{perliger2016gap}. Their motivation could be terrorism as altruism \cite{reeve2020islamist} or for financial reasons to support their family~\cite{krueger2003education}. For these reasons we will refer to the "perpetrator" as the individual who attempts to carry out the terrorist attack. We will create a profile for this actor with parameters as low income, low-medium education, not married, young, highly religious.

The second role would be of the "master mind" of the attack~\cite{russell1977profile}. Normally these people carry a special persona and dedicate their lives to a mission. They are regarded as spiritual leaders. In Islamic terrorist groups, there is evidence that leaders are likely to be married, employed, and found religion later in life~\cite{perliger2016gap}. We simulated agents for our models that we refer to as "leaders"; their profile parameters are low income, medium-high education, married, middle-aged, highly religious.

 Another distinct role we need for a successful attack is financing the operations. These actors in our modeling would have a medium-high income for the environment they are financing or residing in, they would have medium-high education, are mostly married, of various ages, and at times might display medium-high religious beliefs.
 We depict these characteristics in Table~\ref{table:1} to synthesise the connection between a certain terrorist profile and their role. There the agents' characteristics are set based on known distributions. If we are unsure, do not have supporting evidence, or there there are strong conflicting or non-converging opinion in the qualitative studies for a particular characteristic, this feature in the agent profile was not populated. We did not have conclusive data for the financier's ``education-status'', the leader's ``wealth-level'' or the perpetrator's ``religious-training''.

At any point in time the simulation contains agents under five profiles \textit{perpetrators, leaders, financiers, civilians}, and \textit{police}. 
 We make  theoretical assumptions of individual terrorists, which reproduce empirical data on terrorist attacks in qualitative analysis \cite{perliger2016gap}.  

The police agents have the role to prevent and stop the conflict, while the civilians can be recruited to become terrorists or denounce a plot they came in contact with. All connections in the society do not denote proximity, but an ideological affinity.

\paragraph{Profile Implementation.} Each agent has a series of features or parameters as depicted in Table~\ref{table:3}. The corroborated value of these features, a tensor value, determines the state of the agent at a certain time. These features are $-$ (i) education status, (ii) marital status, (iii) wealth level, (iv) religious training, (v) exposure to crime, remain constant over time. The values of these features are sampled from a normal distribution that follows the characteristics of an actual society (i.e. countries, cities, states, a.s.o.) present within certain geographical boundaries.

In addition to these parameters that describe the incipient state of an agent and remain constant throughout the simulation, additional parameters are used to help them adapt and function in the environment. The value of these parameters change throughout the evolution of the model and they are: (vi) crimes committed, (vii) predisposition towards police, (viii) predisposition towards terrorism, and (ix) power level. All agents in the environment have these nine features. Based on the corroboration of their values agents can be assigned to a certain group described above: terrorist (financier, perpetrator, leader), civilian, or police. Their profile (stored in a rank 9 tensor, Table~\ref{table:3}) in the synthetic environment dictates the role they play. The agents are adaptive, therefore the value of the last four characteristics change over time. Subsequently one particular agent can play different roles in the course of a simulation; we define this as the current state of the agent.
These traits were chosen for functionality of the model based on~\cite{perliger2016gap}.

\begin{table}[h!]
\begin{center}
\caption{Hypothesis of Agent Traits}
\label{table:3}
\scriptsize
\begin{tabular}{ccccc}
\toprule
   {Agent Traits} & Civilian & police & Terrorist \\ \hline
   \textit{Education Status} & {constant} & {constant} & {constant} \\
   \textit{Marital Status} & {constant} & {constant} & {constant} \\
   \textit{Wealth Level} & {constant} & {constant} & {constant} \\
   \textit{Religious Training} & {constant} & {constant} & {constant} \\
   {Exposure to Crime} & {constant} & {constant} &{constant} \\
   {Crimes Committed} & \textit{variable} & \textit{variable} & \textit{variable} \\
   {Predisposition towards Police} & \textit{variable} & \textit{variable} & \textit{variable} \\
   {Predisposition towards Terrorism} & \textit{variable} & \textit{variable} & \textit{variable} \\
   {Power} &\textit{variable} & \textit{variable} & \textit{variable} \\
\hline
\end{tabular}
\normalsize
\end{center}
\end{table}


\textbf{Notation.} Traditionally, in ABM environments  "time" is emulated by subsequent runs of the environment referred to as "ticks". Therefore for every tick $t$ an agent $\alpha_i$ will have the state {${\alpha}_i^{t}$}. The overall state of an agent $\alpha$ is stored in a rank 9 tensor. The individual features $\tau$ differentiated as $\tau_1, \tau_2,\tau_3, \tau_4, \tau_5, \tau_6,\tau_7, \tau_8, \tau_9 $,  represent agent traits as listed in (Table 3) in ascending order.

While the environment evolves, certain agent traits remain constant while others change as a result of agent interaction. We present the hypothesised fixed and variable agent traits in (Table~\ref{table:3}).
The constant characteristics ($\tau_1, \tau_2,\tau_3, \tau_4, \tau_5$) are populated at the inception of the environment with $high$, $low$, or $neutral$ using normal distribution without feature correlation (Figure 2(a)). As the world advances (in ticks) the values of node features ($\tau_6, \tau_7,\tau_8, \tau_9 $) can change following simple rules (Algorithm~\ref{algorithm:seeding1}).

 \textbf{Agent Initialization.} This simulation can be represented as an undirected graph, where the nodes are the agents and the edges represent an \emph{ideological connection} between two agents. To initialize the society, we first instantiate the nodes.

 Subsequently a random number of connections are created, to ensure a minimum of two connections for each node, with \textless 10\% of the nodes being part of small random highly connected clusters rather than a complete random graph. This simulates the scattered groups of loyalty and translates to affinity towards people in society (it could be family, or ideology based affinities).  

 All the states of an agent $\alpha_1$ from initial iteration $1$ to final iteration $m$ would be:
 
\begin{center}
\begin{math}
\alpha_1^1=\{ \tau_1,\tau_2,\tau_3,\tau_4, \tau_5,\tau_6^1, \tau_7^1,\tau_8^1, \tau_9^1\}
\end{math}
\BlankLine
\par
\begin{math}
\alpha_1^2=\{ \tau_1,\tau_2,\tau_3,\tau_4, \tau_5,\tau_6^2, \tau_7^2,\tau_8^2, \tau_9^2\}
\end{math}
\\
\begin{math}
\alpha_1^m=\{ \tau_1,\tau_2,\tau_3,\tau_4, \tau_5,\tau_6^m, \tau_7^m,\tau_8^m, \tau_9^m\}
\end{math}

\end{center}

\begin{algorithm}[!htp]
\DontPrintSemicolon
\caption{Constructing the synthetic environment}
\SetKwInOut{Input}{Input}
\SetKwInOut{Output}{Output}
\SetKw{KwRet}{return}
\footnotesize
\Input{Agent feature decomposition:
$\alpha^i=\{ \tau_1,\tau_2,\tau_3,\tau_4, \tau_5,\tau_6^i, \tau_7^i,\tau_8^i, \tau_9^i\}$.
Establish geographical area indicators ; World size; Number of runs.}
\Output{Synthetic data}
\textbf{Method:}\;
\textbf{Step 1.}  Calculate probability density function based on the reported parameters in considered geographical area.\\ Weight := \{education status, marital status, religion training, relative wealth, exposure to crime\}. \;
\textbf{Step 2.}  Call in function to populate the environment with variable trait values: exposure to crime, crimes committed, predisposition towards police, predisposition towards terrorism, power level. \;
\textbf{Step 3.}  Environment setup: generate and balance agent features based on the established distributions for the overall synthetic environment and agent level per feature\;
\textbf{Step 4.}  Ensure ideological links 1:1 and few highly connected groups\;
\textbf{Step 5.}  Write update state after every tick.
\label{algorithm:seeding}
\normalsize
\end{algorithm}

\begin{algorithm}[!ht]
\DontPrintSemicolon
\caption{Adaptive Agents Interaction}
\SetKwInOut{Input}{Input}
\SetKwInOut{Output}{Output}
\SetKw{KwRet}{return}
\footnotesize
\Input{ Agent feature decomposition (see section 3.2):  \begin{math}
            \alpha^i=\{ \tau_1,\tau_2,\tau_3,\tau_4, \tau_5,\tau_6^i, \tau_7^i,\tau_8^i, \tau_9^i\}
            \end{math}; Establish geographical area indicators (see section 3.3); World size; Number of runs.}
\Output{Outcome of an interaction}
\textbf{Method:}\;
\If{agent has a neutral predisposition}{
(a) collides w/\textit{terrorist}, increases terrorist predisposition\\  
(b) collides w/ \textit{police}, increases police predisposition\
}
\If{agent meets threshold for predisposition towards police}{
    (a) collides w/          \textit{terrorist}:\\
    Stage 1: no action takes place. Subsequent interactions result in removal(arrest) \;
    (b) collides w/ \textit{police},  both police increase power \;
    (c) collides w/ \textit{civilian}, police predisposition increases\;
     }

\If{agent meets threshold predisposition towards terrorism}{
    \If{leader}{
      collides w/ \textit{terrorist}, leader power increases \\
      \If{power $>$ threshold}{attack $\implies$ crimes committed increases\;}
      (b) collides w/ \textit{police}, power decreases \tab[0.1cm]and \\
      \If{power $<$ threshold}{ $\implies$ leader removed\;}
      (c) collides w/ \textit{civilian}, terrorist predisposition increases \tab[0.1cm]and \\
      \If{predisposition $>$ threshold}{ $\implies$ recruit\;}
    }
    \eIf{agent meets threshold financier}{
   (a) collides w/terrorist, both increase power\\
   (b) collides w/ \textit{police}, power decreases \tab[0.1cm]and \tab[.1cm]\\
   \If{power drops under a set threshold}{leader removed}
   (c) collides w/ \textit{civilian}, terrorist predisposition increases \\
    }
    {
       collides w/terrorist $\longrightarrow$ terrorist power increases and \\
      \If{power $>$ threshold}{attack $\implies$ committed crimes increases\;}
      (b) collides w/ \textit{police} : power decreases \tab[0.1cm]and\\
       \If{power $<$ threshold}{terrorist removed}
       (c) collides w/ \textit{civilian}, terrorist predisposition increases\;
    }
  }
\label{algorithm:seeding1}
\normalsize
\end{algorithm}

\textbf{Environment setup.} We constructed an uniform environment in a continuous, unbounded space. We instruct an infinite plan topology consistent with ideological loyalty independent of space. 

\textbf{Interactions.} In this environment, interactions are limited to occur between the agents and modeled as agent states $i$. The connections between agents are not to be interpreted as physical proximity, but devotion, affinity, and doctrine.  These ideological connections are modelled by randomly initiated edges and clusters of \textit{loyally}. The interaction outcome is initially modeled as a logistic distribution.

In addition to creating the agents, we model interaction rules for agents to interconnect. We assume the distribution of attributes for each role and compare recorded events of terrorism after each iteration of the process.

 To represent all nodes and their states, we will use additional subscripts to capture the uniqueness of a certain characteristic of a node. For instance, an agent $\alpha_1^\textit{t}$(agent one at time $t$) will have its own predisposition towards terrorism $\tau_{18}^\textit{t}$ reflected at the global environment level as ${\tau}$ with a subscript $_1$ meaning it belongs to agent one followed by subscript $_8$ meaning it is the 8$^{th}$ value of the tensor with superscript $^t$ showing it is captured at tick/time \textit{t}:

\begin{center}
\begin{math}
\alpha_1^\textit{t}=\{ \tau_{11},\tau_{12},\tau_{13},\tau_{14}, \tau_{15},\tau_{16}^\textit{t}, \tau_{17}^\textit{t},\tau_{18}^\textit{t}, \tau_{19}^\textit{t}\}
\end{math}
\BlankLine
\par
\begin{math}
\alpha_2^\textit{t}=\{ \tau_{21},\tau_{22},\tau_{23},\tau_{24}, \tau_{25},\tau_{26}^\textit{t}, \tau_{27}^\textit{t},\tau_{28}^\textit{t}, \tau_{29}^\textit{t}\}
\end{math}
\BlankLine



\par

\par

\begin{math}
\alpha_i^t=\{ \tau_{i1},\tau_{i2},\tau_{i3},\tau_{i4}, \tau_{i5},\tau_{i6}^t, \tau_{i7}^t,\tau_{i8}^t, \tau_{i9}^t\}
\end{math}
\BlankLine
\par

\par
\begin{math}
\alpha_n^t=\{ \tau_{n1},\tau_{n2},\tau_{n3},\tau_{n4}, \tau_{n5},\tau_{n6}^t, \tau_{n7}^t,\tau_{n8}^t, \tau_{n9}^t\}
\end{math}

\end{center}

To model the probability of a loyalty connection, we assume logistic distribution dependant on the profile  polarization of the agents and mean of the modelled distribution.  We use a cumulative distribution function to model the probability of success for each interaction as computed below
 \begin{equation}
 \nonumber
  F(x)=     \frac{1}{1+ \exp\left(\frac{-(\text{w}*\text{$\tau$})- \text{$\mu$})}{s}\right)}, \label{eq:logit}
 \end{equation}
 where we have specific geographic rates ($w$), agent traits ($\tau$), mean distribution ($\mu$), and scale of the logistic distribution ($\textit{s}$). The values for $w$ and $s$ are chosen based on the geographical recorded data for a certain region, where $\mu$ and $\tau$ are dynamically calculated for with each interaction, during the synthetic environment.

\section{Modeling and Simulation}
 For the initial setup
 the following recorded distributions of real data \cite{UN,WorldBank} are considered: (a) percent of married population aged 15 years and older, (b) population wealth index based on Poisson distribution for scaled relative wealth to geo-location, (c) religion distribution, (d) education level (decimal point) ($e_1$ no education, $e_2$ primary,  $e_3$ secondary, $e_4$ tertiary education); the total value of combined education will sum up to one :
$e_1+e_2+e_3+e_4=1$, and (f) crime as exposure and density for considered area.

For calibration, we used Maryland terrorism database~\cite{UMD}. The limitations of this dataset were in recording terrorist activities as following: in order for an incident to be in the database, the action needs to be intentional with violence or threat of violence with non-national actors involved.  It also must have two of the following: be outside of war activity, have an intended message for others beyond the victims, or had a declared a social, religious, political, or economic goal. The data used was from 2012-2017 as the method of collection prior to 2012 was significantly different.

The secondary features for the model tried to introduce demographic indexes on Libya, France, and Pakistan, due to their different profiles for the 9 characteristics considered. 
Educational data was extracted from the World Bank/Barro Lee data-set \cite{WorldBank}. Marital status was found through United Nations Statistical Department \cite{UN}. Religious orientation was modeled using Pew Research Center data\cite{muslim}. Recorded crimes were pulled from UMD database\cite{UMD}.

\subsection{Parameters, Interactions, and Outcome}
 The individual terrorist profile desegregation was used as the hypothesis to support the mathematical formulation. At every point in time the profiles of each agent were recorded as values in a matrix. While there was no specific label for the agents, however the \textit{predisposition towards police or terrorism} parameter value would indicate the behavior of the agent oriented towards  terrorist, civilian, or police(Figure 2(a),(b),(c)), where we notice (on $x-$axis) the predisposition towards terrorist or police as a value that evolves over the course of synthetic environment. In consequence, the synthetic environment was configured and it is briefly described in Algorithm 1. 
   \begin{figure*}[!htp]
         \centering
         \includegraphics[clip,trim=0cm 0cm 0.5cm 0cm, width=0.86\textwidth]{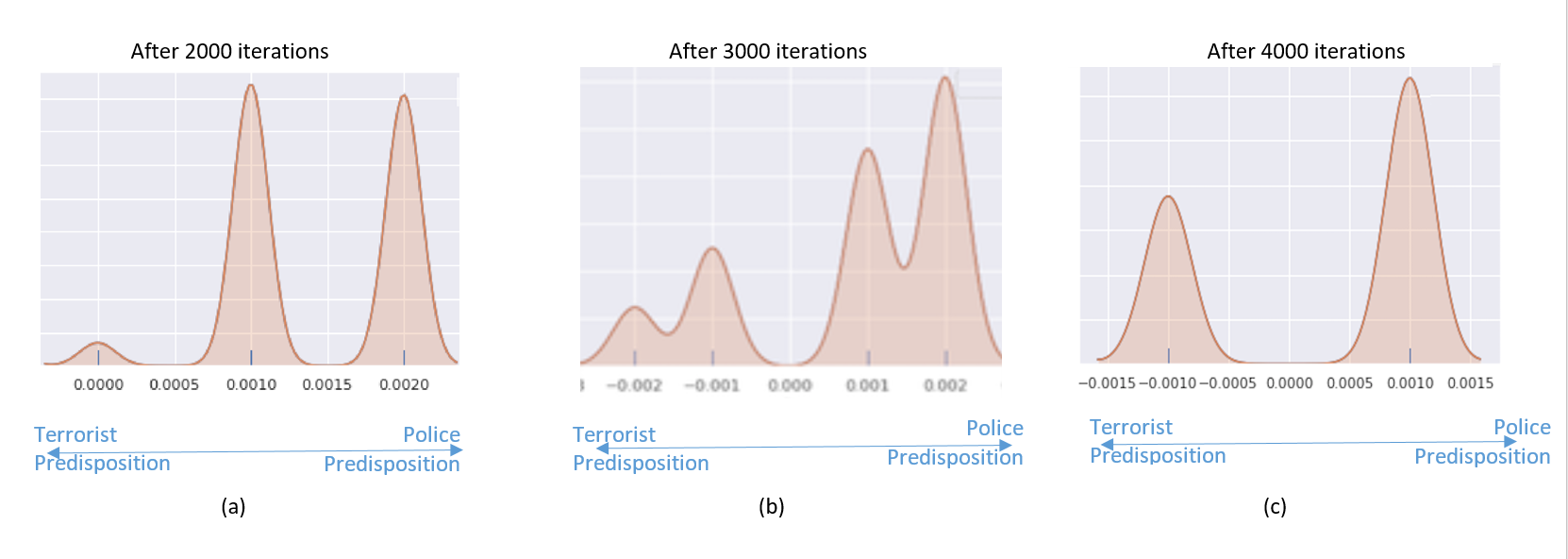}
         \includegraphics[clip,trim=0cm 0cm 0.5cm 0cm, width=0.86\textwidth, keepaspectratio]{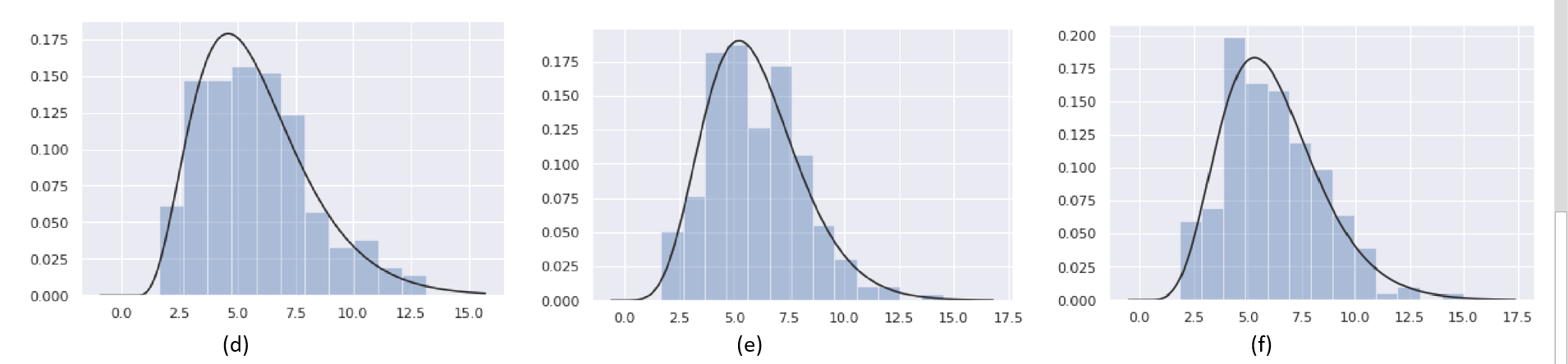}
         \caption{Synthetic environment: 
         (a) after 2000 iterations, society is highly positive, x-axis shows the polarity of the population towards terrorist or law enforcement support. We notice a high positive polarization towards the police and some neutrality
         (b) after 3000 iterations, we notice some very negative attitudes
         (c) after 4000 iterations, the society has a high divide towards police or terrorism (d) consecutively after 2000 iterations, the attacks start to show a long tail towards the right  (e) after 3000 iterations, we notice accentuated long tail to the right, with high variation in values
         (f) after 4000 iterations, we notice long tail rare events and variance, comparable to recorded data Figure 1(d).}
         \label{fig:caseA}
    \end{figure*}
    
 \textit{Interaction outcome} between two nodes (agents) with a high predisposition towards terrorism could constitute the bases of an attack or recruitment into a different role. The probability of success of such an action, will be modeled through a logistic distribution. If the probability from the logistic distribution is greater for a certain role, the individual is recruited to this role.  Whenever an agent has interaction with its own kind their power increases. However, when a police individual meets a terrorist, nothing happens until they exceed a set level of power threshold. Once a leader has at least three underlings and has accumulated enough power, the leader can begin to plan an attack, if a connection with a financier profile of enough power is present. Now if the police personnel interacts with a terrorist, then the terrorist can be arrested. The the success of the attack comes from a logistic distribution based on the power, number of terrorist, and right probabilistic context from the other agents they are connected with. If for any reason an agent is removed from the environment, its place is replaced with a new agent with profile values matching the initial environment setup. A set of simple rules for outcome of an interaction is presented in Algorithm 2.
  
Recorded statistics of the geographical country are embedded into the model. For each simulation, the recorded run results were through cycles of 1000, 2000, 3000, and 4000 iterations of interactions (ticks). Because we found a strong divide between the profiles after 4000 iterations, we did not run the environment further.  

 We extracted the data and considered the medians of these results. We assume terrorist attacks as rare events, so we look for those which have a median of zero compared to those with higher medians. Those with higher medians are having a large number of attacks. We have noticed that Latin hypercube sampling (LHS) yield a better spread than our initial random sampling.
 Figure 2 shows the data visualizations recorded for the predisposition towards terrorism as opposed to predisposition towards police. Predisposition towards terrorism would account for negative values on the $x-$axis, and positive values on this axis would show a favorable attitude, predisposition towards police. Figure 2 (a) shows the predisposition after 2000 runs the population is highly in favor of the police, and there is some neutrality. In Figure 2 (b), we show the predisposition distribution after 3000 runs. Here we notice that there is significantly higher predisposition towards terrorism. The population evolves very fast in the next runs into a strong divide of agent polarization as shown in Figure 2 (c).
For this runs we are showing the types of actions that take place in the synthetic environment. At the beginning, the environment is initialized with a normal distribution of profiles. Figures 2 (d), (e) and (f) show the model evolution over time and we notice the elongation of right-hand tails, which models the terrorist attacks. Over time the events remain very rare in the synthetic environment.

\subsection{Discussion}
This exploratory work at the intersection of political science and modeling in computer science, takes the theoretical research of rare events, one step further by implementing interdisciplinary concepts. The simulation of an entire society including terrorist profiles as an unbounded network linked by affinity, showed initial evidence towards producing pertinent synthetic data for mission critical applications.

Further investigations are needed to fine-tune the role of each variable. So far we validated the results under different assumptions using various modalities of sampling. We notice a slight predisposition preference for the weights  in our initial hypothesis and chanced to employ LHS. We compared the mean and variance distribution of data extracted from the synthetic distribution with the data from real events. For deadly events in the RAND-MIPT data-set we record a low mean of 4.2 and a very high variance of 727. This matches our synthetic environment mean and variance just before 4000 ticks (mean 3.1, variance 748), when the population becomes strongly divided. The model's convergence or not, in this case, does not speak for an agent-based-model's competence, but for its fine tuning.

\section{Conclusion}
 In this manuscript we propose a novel approach for generating synthetic data that triggers anomaly detection in complex social networks. This interdisciplinary research brings political-social innovation into data sciences;  bridging mature qualitative research from other areas of science into computer science modelling. To do so, a set of societal rules were used to prototype an autonomous multi-agent synthetic environment with self-organized terrorist behavior within the context of a whole society. Subsequently, we prototype extraction of long tailed synthetic data-set that showed basic measures of similar mathematical distributions to collected data~\cite{asmussen2000rare}: density, mean, and variance. This methodology could serve as a guideline to develop more sophisticated mathematical rigors towards data generation for practical machine learning with strict limitation or lack of sufficient recorded data.
 We have also identified some challenges and propose future research directions.

Firstly we want to focus on the sensitivity condition. The model is sensitive to the primary initial conditions dealing with  adaptive intelligent agents profiles. Evidence of sensitivity of our dis-aggregated profiles behavior is shown in the graph 2(c) where the society becomes segregated. However the model does not show substantially sensitive to our subsequent (nice-to-have) geographical conditions.  Since the geographical conditions are a secondary, independent set of restrictions, we believe fine tuning of the model would address these.
With regards to the number of iterations, we opine that the model shows the right behavior. At about 4000 ticks it becomes highly segregated and if we continue the runs (which we do not show in this paper) it continues this pattern we were expecting to see. This shows our agents are eager to maximize their objective function.

While proving a valid initial proof of concept for desegregation of rare events given a set of premises we notice the necessity of a visual environment to allow fine tuning of the conditions, as opposed to only measuring the mathematics. To this end, we propose future work implementation of our algorithm in NetLogo. NetLogo is a free, multi-agent programmable modeling environment, developed by Uri Wilensky, at The Center for Connected Learning and Computer-Based Modeling, at Northwestern University\cite{wilensky2015introduction}. This interactive programming environment accounts for the ability to fine tune input parameters and visually observe the effect it has on the synthetic society. Here we would aim for a model that does not converge, allowing the world to keep on going, with continuous rare events taking place. A NetLogo implementation would visually show how each condition corroborates in the overall environment and help identify optimal tuning. 
 
 Secondly, we propose introducing some factors that would model the probability of Perceived Mission Risk of Detection and enforcing the reward (Ilachinski, 2012). We propose stricter identification of distance between input variants and output distributions in the temporal context of runs. As a future direction we propose expanding the focus to generating synthetic data in this domain that matches complex overall distributions, and follows more rigorous mathematical properties. One of the metrics KL divergence, in addition to checking for co-variance structure, outliers, etc.).

Lastly we think future work research would benefit from introduction of time varying constrains in relation to objectives and parameters for more practical optimization problem as it relates to the evolution of agents in the synthetic environment~\cite{6900487}.
 
 \BlankLine
 \BlankLine

\section{Acknowledgement}
We acknowledge \textbf{Ms. Melody Walker} from Virginia Tech Mathematics Department contributions as co-author towards research, producing the code, and partial results during her 2019 NSF appointment with GRL US Army.  
 \textbf{Dr. Brian Jalaian} from CCDC Army Research Laboratory for very specific questions,  technical recommendation. His input was vital to clearly address the impact of current research. We are deeply appreciative for  
 \textbf{Dr. Jerry Ballard}, \textbf{Dr. Brian Jalaian}, and \textbf{Prof. Naren Ramakrishnan} ongoing support in advancing the work. {Ms. Nicole Wayant, Ms. Charlotte Ellison}, GRL US Army, and Prof. Mihai Boicu, George Mason University for reviewing the manuscript. 

This report was prepared as an account of work sponsored by an agency of the United States Government. Neither the United States Government nor any agency thereof, nor any of their employees, nor any of their contractors, subcontractors, or their employees, makes any warranty, express or implied, or assumes any legal liability or responsibility for the accuracy, completeness, or any third party’s use or the results of such use of any information, apparatus, product, or process disclosed, or represents that its use would not infringe privately owned rights. Reference herein to any specific commercial product, process, or service by trade name, trademark, manufacturer, or otherwise, does not necessarily constitute or imply its endorsement, recommendation, or favoring by the United States Government or any agency thereof or its contractors or subcontractors. The views and opinions of authors expressed herein do not necessarily state or reflect those of the United States Government or any agency thereof.

\bibliographystyle{aaai}
{\fontsize{9.0pt}{10.0pt}\selectfont
\bibliography{references}

\begin{thebibliography}{}

\bibitem[\protect\citeauthoryear{Asmussen \bgroup et al\mbox.\egroup
  }{2000}]{asmussen2000rare}
Asmussen, S.; Binswanger, K.; H{\o}jgaard, B.; et~al.
\newblock 2000.
\newblock Rare events simulation for heavy-tailed distributions.
\newblock {\em Bernoulli} 6(2):303--322.

\bibitem[\protect\citeauthoryear{{Biswas} \bgroup et al\mbox.\egroup
  }{2014}]{6900487}
{Biswas}, S.; {Das}, S.; {Suganthan}, P.~N.; and {Coello}, C. A.~C.
\newblock 2014.
\newblock Evolutionary multiobjective optimization in dynamic environments: A
  set of novel benchmark functions.
\newblock In {\em 2014 IEEE Congress on Evolutionary Computation (CEC)},
  3192--3199.

\bibitem[\protect\citeauthoryear{Cedeno-Mieles \bgroup et al\mbox.\egroup
  }{2019}]{cedeno:2019}
Cedeno-Mieles, V.; Hu, Z.; Deng, X.; Ren, Y.; Adiga, A.; Barrett, C.;
  Ekanayake, S.; Korkmaz, G.; Kuhlman, C.~J.; Machi, D.; Marathe, M.~V.; Ravi,
  S.~S.; Goode, B.~J.; Ramakrishnan, N.; Saraf, P.; Self, N.; Contractor, N.;
  Epstein, J.~M.; and Macy, M.~W.
\newblock 2019.
\newblock Mechanistic and data-driven agent-based models to explain human
  behavior in online networked group anagram games.
\newblock In {\em Proceedings of the 2019 IEEE/ACM International Conference on
  Advances in Social Networks Analysis and Mining}, ASONAM ’19,  357–364.
\newblock New York, NY, USA: Association for Computing Machinery.

\bibitem[\protect\citeauthoryear{Center}{2010}]{muslim}
Center, P.~R.
\newblock 2010.
\newblock Table: Muslim population by country.

\bibitem[\protect\citeauthoryear{Chang, Chen, and
  Fomby}{2017}]{ChangChenFomby2017}
Chang, K.; Chen, R.; and Fomby, T.~B.
\newblock 2017.
\newblock Prediction-based adaptive compositional model for seasonal time
  series analysis: Prediction-based adaptive compositional model for seasonal
  time series analysis.
\newblock {\em Journal of Forecasting} 36(7).

\bibitem[\protect\citeauthoryear{Clauset}{2015}]{Clauset2012}
Clauset, A.
\newblock 2015.
\newblock {\em Estimating the probability of rare events}.

\bibitem[\protect\citeauthoryear{Corporation}{2018}]{RAND2018}
Corporation, R.
\newblock 2018.
\newblock {\em Cost of Terrorism}.

\bibitem[\protect\citeauthoryear{Department}{2006}]{UN}
Department, U. N.~S.
\newblock 2006.
\newblock Population by marital status, age, sex, and urban/rural residence.

\bibitem[\protect\citeauthoryear{Galarnyk}{2019}]{galarnyk2019}
Galarnyk, M.
\newblock 2019.
\newblock Explaining the 68-95-99.7 rule for a normal distribution.

\bibitem[\protect\citeauthoryear{Gilbert}{2019}]{gilbert2019agent}
Gilbert, N.
\newblock 2019.
\newblock {\em Agent-based models}, volume 153.
\newblock Sage Publications, Incorporated.

\bibitem[\protect\citeauthoryear{Grimm \bgroup et al\mbox.\egroup
  }{2006}]{grimm2006standard}
Grimm, V.; Berger, U.; Bastiansen, F.; Eliassen, S.; Ginot, V.; Giske, J.;
  Goss-Custard, J.; Grand, T.; Heinz, S.~K.; Huse, G.; et~al.
\newblock 2006.
\newblock A standard protocol for describing individual-based and agent-based
  models.
\newblock {\em Ecological modelling} 198(1-2):115--126.

\bibitem[\protect\citeauthoryear{Ilachinski}{2012}]{ilachinski2012modelling}
Ilachinski, A.
\newblock 2012.
\newblock Modelling insurgent and terrorist networks as self-organised complex
  adaptive systems.
\newblock {\em International Journal of Parallel, Emergent and Distributed
  Systems} 27(1):45--77.

\bibitem[\protect\citeauthoryear{Keller, Desouza, and
  Lin}{2010}]{keller2010dismantling}
Keller, J.~P.; Desouza, K.~C.; and Lin, Y.
\newblock 2010.
\newblock Dismantling terrorist networks: Evaluating strategic options using
  agent-based modeling.
\newblock {\em Technological Forecasting and Social Change} 77(7):1014--1036.

\bibitem[\protect\citeauthoryear{Krueger and
  Male{\v{c}}kov{\'a}}{2003}]{krueger2003education}
Krueger, A.~B., and Male{\v{c}}kov{\'a}, J.
\newblock 2003.
\newblock Education, poverty and terrorism: Is there a causal connection?
\newblock {\em Journal of Economic perspectives} 17(4):119--144.

\bibitem[\protect\citeauthoryear{Li \bgroup et al\mbox.\egroup
  }{2015}]{li2015agent}
Li, B.; Sun, D.; Zhu, R.; and Li, Z.
\newblock 2015.
\newblock Agent based modeling on organizational dynamics of terrorist network.
\newblock {\em Discrete Dynamics in Nature and Society} 2015.

\bibitem[\protect\citeauthoryear{Macal and North}{2005}]{macal2005tutorial}
Macal, C.~M., and North, M.~J.
\newblock 2005.
\newblock Tutorial on agent-based modeling and simulation.
\newblock In {\em Proceedings of the Winter Simulation Conference, 2005.},
  14--pp.
\newblock IEEE.

\bibitem[\protect\citeauthoryear{Macy and Willer}{2002}]{macy2002factors}
Macy, M.~W., and Willer, R.
\newblock 2002.
\newblock From factors to actors: Computational sociology and agent-based
  modeling.
\newblock {\em Annual review of sociology} 28(1):143--166.

\bibitem[\protect\citeauthoryear{Malleson and
  Andresen}{2015}]{malleson2015impact}
Malleson, N., and Andresen, M.~A.
\newblock 2015.
\newblock The impact of using social media data in crime rate calculations:
  shifting hot spots and changing spatial patterns.
\newblock {\em Cartography and Geographic Information Science} 42(2):112--121.

\bibitem[\protect\citeauthoryear{Mullins}{2009}]{mullins2009parallels}
Mullins, S.
\newblock 2009.
\newblock Parallels between crime and terrorism: A social psychological
  perspective.
\newblock {\em Studies in Conflict \& Terrorism} 32(9):811--830.

\bibitem[\protect\citeauthoryear{Perliger, Koehler-Derrick, and
  Pedahzur}{2016}]{perliger2016gap}
Perliger, A.; Koehler-Derrick, G.; and Pedahzur, A.
\newblock 2016.
\newblock The gap between participation and violence: Why we need to
  disaggregate terrorist ‘profiles’.
\newblock {\em International Studies Quarterly} 60(2):220--229.

\bibitem[\protect\citeauthoryear{Perry}{2013}]{perry2013predictive}
Perry, W.~L.
\newblock 2013.
\newblock {\em Predictive policing: The role of crime forecasting in law
  enforcement operations}.
\newblock Rand Corporation.

\bibitem[\protect\citeauthoryear{Reeve}{2020}]{reeve2020islamist}
Reeve, Z.
\newblock 2020.
\newblock Islamist terrorism as parochial altruism.
\newblock {\em Terrorism and political violence} 32(1):38--56.

\bibitem[\protect\citeauthoryear{Ritchie \bgroup et al\mbox.\egroup
  }{2013}]{ritchiehasellappelroser2013}
Ritchie, H.; Hasell, J.; Appel, C.; and Roser, M.
\newblock 2013.
\newblock Terrorism.
\newblock {\em Our World in Data}.

\bibitem[\protect\citeauthoryear{Russell and Miller}{1977}]{russell1977profile}
Russell, C.~A., and Miller, B.~H.
\newblock 1977.
\newblock Profile of a terrorist.
\newblock {\em Studies in conflict \& terrorism} 1(1):17--34.

\bibitem[\protect\citeauthoryear{Singh, Vainchtein, and
  Weiss}{2009}]{singh2009schelling}
Singh, A.; Vainchtein, D.; and Weiss, H.
\newblock 2009.
\newblock Schelling’s segregation model: Parameters, scaling, and
  aggregation.
\newblock {\em Demographic Research} 21:341--366.

\bibitem[\protect\citeauthoryear{START}{2018}]{UMD}
START.
\newblock 2018.
\newblock Global terrorism database.

\bibitem[\protect\citeauthoryear{Tecuci \bgroup et al\mbox.\egroup
  }{1999}]{tecuci1999integrated}
Tecuci, G.; Boicu, M.; Wright, K.; Lee, S.~W.; Marcu, D.; and Bowman, M.
\newblock 1999.
\newblock An integrated shell and methodology for rapid development of
  knowledge-based agents.
\newblock In {\em AAAI/IAAI},  250--257.

\bibitem[\protect\citeauthoryear{Tsvetovat and Carley}{}]{tsvetovat2002knowing}
Tsvetovat, M., and Carley, K.
\newblock {\em Knowing the enemy: A simulation of terrorist organizations and
  counter-terrorism strategies}.

\bibitem[\protect\citeauthoryear{Wilensky and
  Rand}{2015}]{wilensky2015introduction}
Wilensky, U., and Rand, W.
\newblock 2015.
\newblock {\em An introduction to agent-based modeling: modeling natural,
  social, and engineered complex systems with NetLogo}.
\newblock Mit Press.

\bibitem[\protect\citeauthoryear{World~Bank}{2018}]{WorldBank}
World~Bank, W. D.~I.
\newblock 2018.
\newblock Educational dataset barro-lee dataset.

\end{thebibliography}
\par}
\end{document}